\documentclass[twocolumn,twoside,english]{IEEEtran2e}

\usepackage{amsmath}   

\usepackage{amssymb}   
\usepackage{amscd}
\usepackage{babel}
\usepackage{latexsym}  
\usepackage{epsfig}
\usepackage{bbm}

\newtheorem{defi}{Definition}
\newtheorem{lemma}[defi]{Lemma}
\newtheorem{satz}[defi]{Theorem}

\newtheorem{bem}[defi]{Remark}

\newtheorem{exempel}[defi]{Example}

\newcommand{\qed}{\hfill $\blacksquare$}
\newcommand{\tr}{{\operatorname{Tr}\,}}

\newcommand{\C}{{\mathbb{C}}}

\newcommand{\alg}[1]{{\mathfrak{#1}}}
\newcommand{\fset}[1]{{\mathcal{#1}}}

\newcommand{\1}{{\mathbbm{1}}}

\begin{document}

\title{The Capacity of the\\ Quantum Multiple Access Channel}
\author{Andreas Winter\thanks{The author is with Fakult\"at f\"ur Mathematik,
Universit\"at Bielefeld, Postfach 100131, 33501 Bielefeld, Germany.
Electronic address: \texttt{winter@mathematik.uni-bielefeld.de}. Research
supported by the Deutsche Forschungsgemeinschaft via SFB 343
``Diskrete Strukturen in der Mathematik''.}
}

\maketitle


\begin{abstract}
  We define classical--quantum multiway channels
  for transmission of classical
  information, after recent work by Allahverdyan and Saakian.
  Bounds on the capacity region are derived in a uniform way, 
  which are analogous to the classically known ones, simply
  replacing Shannon entropy with von Neumann entropy.
  For the single receiver case (multiple access channel)
  the exact capacity region is determined.
  These results are applied to the case of noisy channels,
  with arbitrary input signal states.
  \par
  A second issue of this work is the presentation
  of a calculus of quantum information quantities,
  based on the algebraic formulation of quantum theory.
\end{abstract}

\section{Introduction}
  \label{sec:intro}
  Classical multiway channels were already studied by Shannon~\cite{shannon:MAC}.
  Ahlswede~\cite{ahlswede:MAC,ahlswede:MWC} first determined the capacity
  region of the channel with $s$ senders and $r$ receivers,
  where all senders want to transmit independent messages,
  which all receivers should get.
  For a good overview on multiuser communication
  theory in general consult~\cite{elgamal:cover}, or
  the textbook~\cite{csiszar:koerner}.
  \par
  In the present paper we define the corresponding quantum channel (after
  Allahverdyan and Saakian~\cite{allahverdyan:saakian}), extending
  the definition of a classical--quantum channel (see~\cite{holevo:channels}).
  Our motivation is twofold: in the first place, it is a very common situation
  that many users want to communicate via the same transmission
  system, and all real systems should be described by quantum mechanics.
  Then, secondly, we feel that it helps understanding quantum communication
  if we try to solve questions known in a classical context for quantum
  channels. This the more, as for Holevo's results on quantum channels
  (coding theorem and information bound) not only the question
  and its answer, but even the method of solution is rather close
  to classically well known mathematics (see~\cite{winter:qstrong}),
  and we should find out whether this similarity extends further.
  \par
  The results of the present work are:
  we bound the capacity region, the
  actual bounds being obtainable from the classical case by formally replacing Shannon
  entropy by von Neumann entropy in the expressions, thus following a general principle
  or feeling in physics. The central result is a proof of the direct coding theorem
  for the multiple access channel (one receiver: $r=1$), using
  the technique of Holevo~\cite{holevo:capacity}
  and Schumacher/Westmoreland~\cite{schumacher:capacity},
  which was designed to solve the single--sender case.
  \par
  The outline of the paper is as follows:
  in section~\ref{sec:channel} the basic definitions are stated,
  in particular quantum multiway channels are formally introduced.
  Section~\ref{sec:basics} reviews notation and facts about quantum
  information quantities we shall need.
  In the following section~\ref{sec:upperbounds} we prove the outer bounds
  for the capacity region.
  Sections~\ref{sec:distortion:fidelity} contains a central result on
  the state disturbance of a measurement with high success probability.
  In section~\ref{sec:directcoding} this result is used to
  prove the direct coding theorem for the quantum
  multiple access channel.
  In the last section~\ref{sec:qq:dmc} we comment
  on the quantum--quantum multiway channel which
  may be fed with arbitrary input states.
  \par
  The results of the present work are part of the author's Ph.D.
  thesis~\cite{winter:qdms}, mainly chapter {III} (with
  alternative proofs), and appendix A.

\section{Quantum Multiway Channels}
  \label{sec:channel}
  This is the simplest situation of multi--user communication in general:
  consider $s$ independent senders, sender $i$ using a (finite)
  alphabet $\fset{X}_i$, say with an
  a priori probability distribution $P_i$. This alphabet serves as a set of
  tags of different actions each user may take, such that a signal
  appears in the output system, composed of the effects of these
  $s$ independent choices, and the channel noise.
  To the output system the $r$ receivers have partial access, and their task
  is to each reconstruct the $s$ messages the senders chose to send.
  This is to be achieved by block--coding and via a previously agreed
  coding/decoding scheme.
  \par
  Formally, this model is captured as follows:
  the channel is simply a map
  \[W:\fset{X}_1\times\cdots\times\fset{X}_s\rightarrow\alg{S}(\alg{Y})\]
  from the input alphabets into the set $\alg{S}(\alg{Y})$ of states of the
  (finite dimensional) C${}^*$--algebra $\alg{Y}$, mapping the
  input $(x_1,\dots,x_s)$ to the output state $W_{x_1\dots x_s}$.
  Without loss of generality we may assume that $\alg{Y}=\alg{L}({\cal H})$
  is the full operator algebra of the finite dimensional Hilbert space
  ${\cal H}$, and we shall assume that the
  $W_{x_1\dots x_s}$ are density operators on
  ${\cal H}$.\footnote{In the general case we may make use of the
    fact that the states
    of $\alg{Y}$ are uniquely described by density operators
    \emph{inside} $\alg{Y}$.}
  \par
  However, to express the theory in this algebraic manner has its merits, as
  we shall see:
  \par
  The output state (generally mixed) is accessed by several receicers.
  These are represented by \emph{commuting} $*$--subalgebras $\alg{Y}_j$
  ($j=1,\ldots,r$) of $\alg{Y}$: the meaning is that receiver $j$
  may use any measurement (POVM) whose elements belong to $\alg{Y}_j$.
  The commutativity ensures that these measurements can be performed
  together. The typical case of this situation is that
  $\alg{Y}=\alg{Y}_1\otimes\cdots\otimes\alg{Y}_r$,
  where we identify $\alg{Y}_j$ with the subalgebra
  $\1^{\otimes(j-1)}\otimes\alg{Y}_j\otimes\1^{\otimes(r-j)}$ of $\alg{Y}$.
  By linear extension we
  may view $W$ as a completely positive, trace preserving map from
  $\alg{X}_{1}\otimes\cdots\otimes\alg{X}_{s}$ to $\alg{Y}$,
  where $\alg{X}_i=\C\fset{X}_i$ is the commutative algebra
  of $\C$--valued functions on $\fset{X}_i$, whose elements
  we identify with their indicator functions.\footnote{Since
    all algebras here are finite dimensional we do not care about
    the topological distictions necessary in general, between
    linear spaces and their duals, between maps and their
    adjoints.}
  If all the $W_{x_1\ldots x_s}$ commute with each other
  (hence have a common diagonalization) the channel is called
  \emph{quasi--classical}, and \emph{classical} if $\alg{Y}$
  is a commutative algebra.
  \par
  It should be stressed that all this can be embedded into
  standard quantum theory by identifying all algebras in
  question with operator algebras in some sufficiently
  large Hilbert space, e.g. a commutative algebra with
  a set of diagonal matrices. To get more familiar with
  this formalism the reader might consult a book
  like~\cite{ohya:petz}.
  \par
  For fixed a priori distributions define the \emph{channel state}
  $$\gamma=\sum_{\forall i\ x_i\in\fset{X}_i} P_1(x_1)\cdots P_s(x_s)
     x_1\otimes\cdots\otimes x_s\otimes W_{x_1\ldots x_s}$$
  on $\alg{X}_1\otimes\cdots\otimes\alg{X}_s\otimes\alg{Y}$.
  This serves as the quantum analogue of the joint distribution
  of the random variables representing input
  and output letters in the classical case. It may be interpreted as
  the joint state of the system after the channel usage, where the
  senders kept a record of their individual letters (this is possible
  because they input classical information, reflected in the
  classical nature of their systems).
  \par
  Note the 1--1--correspondence between states $\gamma$
  on $\alg{X}_1\otimes\cdots\otimes\alg{X}_s\otimes\alg{Y}$
  and pairs $(P,W)$ of channels $W$ and probability
  distributions $P$ on $\fset{X}_1\times\cdots\times\fset{X}_s$.
  This is a feature of our model, which relies on the
  commutativity of the $\alg{X}_i$ (compare~\cite{ohya:qchannels}
  for the difficulties encountered in more general situations).
  \par
  We will employ block coding on the discrete memoryless channel
  generated by $W$: for sequences
  $x_1^n,\ldots,x_s^n$, ${x}_i^n=x_{i1}\ldots x_{in}\in{\cal X}_i^n$,
  the $n$--block channel
  $$W^n:{\cal X}_1^n\times\cdots\times{\cal X}_s^n
                           \rightarrow\alg{S}(\alg{Y}^{\otimes n})$$
  is defined by
  $$W^n_{{x}_1^n\ldots{x}_s^n}\!=\! W_{x_{11}x_{21}\ldots x_{s1}}\!\otimes\!
                                    W_{x_{12}x_{22}\ldots x_{s2}}\!\otimes\!\cdots
                        \!\otimes\! W_{x_{1n}x_{2n}\ldots x_{sn}}.$$
  We now introduce some notation to describe the channel as seen by
  a subset of senders, while the others enter only stochastically:
  \par
  For a $J\subset[s]=\{1,\ldots,s\}$ denote
  $P_J=\bigotimes_{i\in J} P_i$, i.e.
  $P_J(x_i|i\in J)=\prod_{i\in J} P_i(x_i)$, and
  ${\cal X}(J)=\prod_{i\in J}{\cal X}_i$ (similarly
  $\alg{X}(J)=\bigotimes_{i\in J}\alg{X}_i$).
  \par
  Define the \emph{reduced channel}
  $P_{J^c}W:{\cal X}(J)\rightarrow\alg{S}(\alg{Y})$ by
  $$(P_{J^c}W)_{(x_i|i\in J)}=
      \sum_{\forall i\in J^c:\ x_i\in\fset{X}_i} P_{J^c}(x_i|i\in J^c)W_{x_1\ldots x_s}.$$
  (Here $J^c$ denotes the complement $[s]\setminus J$ of $J$ in $[s]$).
  Note that
  $$\tr_{\alg{X}(J^c)}\gamma=\!\!\sum_{\forall i\in J:\ x_i\in\fset{X}_i}\!\!
                         P_J(x_i|i\in J)\bigotimes_{i\in J} x_i
                                         \!\otimes\! (P_{J^c}W)_{(x_i|i\in J)}.$$
  Transmission is now by using codes on $n$--blocks:
  \par
  An $n$--\emph{block--code} is a collection
  $(f_1,\ldots,f_s,D_1,\ldots,D_r)$ of maps
  $f_i:\fset{M}_i\rightarrow\fset{X}_i^n$ (where $\fset{M}_i$ is the set of
  messages of sender $i$) and decoding observables (POVMs)
  $D_j\subset\alg{Y}_j^{\otimes n}$, indexed by
  $\fset{M}_1\times\cdots\times\fset{M}_s$, i.e.:
  $$D_j=\{D_{j\mu}\in\alg{Y}_j:\mu\in\fset{M}_1\times\cdots\times\fset{M}_s\},$$
  such that
  $$D_{j\mu}\geq 0,\quad \sum_{\mu} D_{j\mu}=\1.$$
  There are $r$ (average) \emph{error probabilities} 
  of the code, the probability that the receiver $j$ guesses incorrectly
  any one of the sent words, taken over the uniform distribution on the
  message sets:
  \begin{equation*}\begin{split}
    \bar{e}_j &(f_1,\ldots,f_s,D_j)= \\
              &1\!-\!\frac{1}{|\fset{M}_1|\cdots|\fset{M}_s|}
                \!\sum_{\forall i:m_i\in\fset{M}_i}
                     \!\!\!\tr\!\left({W^n_{f(m_1)\ldots f(m_s)}
                                      D_{j,m_1\ldots m_s}}\right)\!.
  \end{split}\end{equation*}
  We call $(f_1,\ldots,f_s,D_1,\ldots,D_r)$ an $(n,\bar{\epsilon})$--code
  if all error probabilities $\bar{e}_j(f_1,\ldots,f_s,D_j)$ do not
  exceed $\bar{\epsilon}$.
  \par
  The \emph{rates} of the code are the $R_i=\frac{1}{n}\log|\fset{M}_i|$.
  A tuple $(R_1,\ldots,R_s)$ is said to be \emph{achievable}, if for any
  $\bar{\epsilon},\delta>0$ there
  exists for any large enough $n$ an $(n,\bar{\epsilon})$--code
  with $i$--th rate at least $R_i-\delta$.
  The set of all achievable tuples (which is clearly closed) is called the
  \emph{capacity region} of the channel, and to determine this region is
  the problem to be addressed here.
  \par
  Some observations should be made: first, the capacity region is
  \emph{convex}, by the \emph{time sharing principle}: 
  let $(R_1,\ldots,R_s)$ and $(R_1',\ldots,R_s')$ be rate
  tuples of $m$-- and $n$--block codes, respectively, with error
  probability $\bar{\epsilon}$ each.
  By concatenating the codewords to $(m+n)$--blocks, and tensoring
  the corresponding decoding observables, we get an $(m+n)$--block
  code with error probability at most $2\bar{\epsilon}$, and with
  rates $\frac{m}{m+n}R_i+\frac{n}{m+n}R_i'$.
  \par
  Second, note that in the multi--user situation
  not a single number describes the performance of the channel
  (as with capacity in the single--sender case). Instead, only
  with given behaviour of the other senders the channel gets
  a specific capacity for a particular sender. Intuitively, this
  is because the others' (unknown!) actions may be seen as
  additional noise (a phenomenon known as ``interference'' in
  classical multi--user channels).

\section{Information Quantities in Quantum Systems}
  \label{sec:basics}
  In this section we introduce some notation in which we express
  our results. 
  From~\cite{winter:qdms}, appendix A,
  we use the definitions of various information
  quantities for observables and $*$--subalgebras, which we review for
  the sake of self--containedness:
  \par
  Let $\alg{A}$ be a C${}^*$--algebra, and $\rho$ a state on it.
  For a $*$--subalgebra $\alg{B}$ we want to define the
  \emph{entropy of $\rho$ with respect to this subalgebra}
  (we shall stress the dependence on $\alg{B}$, as $\rho$ is supposed
  to be fixed). To this end let us consider the restriction
  $\sigma=\rho|_{\alg{B}}$ of
  $\rho$ to $\alg{B}$, and define
  $$H(\alg{B})=H_\rho(\alg{B})=-\tr(\sigma\log\sigma).$$
  Here $\tr$ is the unique \emph{trace} on $\alg{B}$ (i.e. a positive
  $\C$--linear functional on $\alg{B}$, with the properties
  $\tr AB=\tr BA$ and $\tr A^*=\overline{\tr A}$), that
  assigns $1$ to all minimal idempotents of $\alg{B}$.
  An important example is the usual trace of $\alg{L}({\cal H})$,
  in which case the formula gives the familiar
  von Neumann entropy of the state.
  \par
  Motivated by identities for classical Shannon entropy we may now define,
  for (elementwise) commuting $*$--subalgebras $\alg{B}$ and $\alg{C}$:
  the conditional entropy
  $$H(\alg{B}|\alg{C})=H(\alg{BC})-H(\alg{C}),$$
  and the mutual information
  \begin{equation*}\begin{split}
    I(\alg{B}\wedge\alg{C}) &=H(\alg{B})+H(\alg{C})-H(\alg{BC}) \\
                            &=H(\alg{B})-H(\alg{B}|\alg{C}).
  \end{split}\end{equation*}
  The condition that the algebras commute is crucial here: it ensures
  that all observables in $\alg{B}$ are coexistent with all observables
  in $\alg{C}$, and also, that the product $\alg{BC}$ is indeed the
  algebra generated by $\alg{B}$ and $\alg{C}$. Of course, these
  definitions are only formally derived from well known classical
  formulas, and there is no reason to expect that they are meaningful
  (which indeed they are only to a limited degree: see the discussions
  in~\cite{cerf:adami}, and in~\cite{levitin}). Anyhow, for
  our purposes they make sufficient sense.
  \par
  If $\alg{D}$ is a third $*$--subalgebra, commuting with both
  $\alg{B}$ and $\alg{C}$, we may finally define the
  conditional mutual information
  \begin{equation*}\begin{split}
    I(\alg{B}\wedge\alg{C}|\alg{D}) &= H(\alg{B}|\alg{D})+H(\alg{C}|\alg{D})
                                      -H(\alg{BC}|\alg{D})                  \\
                                    &= H(\alg{BD})+H(\alg{CD})
                                      -H(\alg{BCD})-H(\alg{D}).
  \end{split}\end{equation*}
  We note, that the conditional mutual information is
  positive, by the strong subadditivity of von Neumann entropy
  (see~\cite{winter:qdms}, theorem~A.9).
  \par
  In all the above expressions we supressed the dependence on the
  underlying state $\rho$. In cases of possible ambiguity it
  is added as a subscript.
  \par
  With these definitions we have the (easily checked) identities for
  the system introduced in section~\ref{sec:channel},
  with the channel state $\gamma$:
  \begin{align*} 
    I(\alg{X}(J)\wedge\alg{Y}|\alg{X}(J^c))
                         &= I(\alg{X}(J)\wedge\alg{Y}\alg{X}(J^c))            \\
                         &= H(\alg{Y}|\alg{X}(J^c))-H(\alg{Y}|\alg{X}_1\cdots\alg{X}_s)
                                                                              \\
                         &= H(P_J W|P_{J^c})-H(W|P_{[s]}),
  \end{align*}
  where in the last line an alternative notation is used:
  \par
  For a channel
  $V:\fset{A}\rightarrow \alg{S}(\alg{Z})$ and a probability distribution
  $Q$ on $\fset{A}$ let
  $$H(V|Q)=\sum_{a\in\fset{A}} Q(a)H(V_a),$$
  with the von Neumann entropy $H$: so this is the familiar
  writing of a conditional as an average of entropies.
  \par
  There are a number of important relations between all these
  quantities, of which we shall make use of two:
  \begin{lemma}
    \label{lemma:subadd}
    Let $V_k:\fset{A}_k\rightarrow \alg{S}(\alg{Z}_k)$ ($k=1,2$)
    be two channels, and $Q$ a probability distribution on
    $\fset{A}_1\times\fset{A}_2$. Forming the channel state
    $$\gamma=\sum_{a_k\in\fset{A}_k:k=1,2}
                    Q(a_1,a_2) a_1\otimes V_{1,a_1} \otimes a_2\otimes V_{2,a_2},$$
    we have the following subadditivity relation:
    $$I(\C\fset{A}_1\C\fset{A}_2\wedge \alg{Z}_1\alg{Z}_2)
        \leq I(\C\fset{A}_1\wedge \alg{Z}_1)+I(\C\fset{A}_2\wedge \alg{Z}_2).$$
  \end{lemma}
  \begin{proof}
    This is well known for classical channels, and the proof in our
    case runs exactly the same. Compare~\cite{winter:qdms},
    theorem~A.17.
  \end{proof}
  \begin{lemma}[Fano inequality]
    \label{lemma:fano}
    Let $\alg{X}$, $\alg{Y}$ be commuting algebras, and
    $\alg{X}$ be commutative. For a state
    $\rho$ on $\alg{XY}$ consider POVMs $X\subset\alg{X}$,
    $Y\subset\alg{Y}$, running over the same index set.
    \par
    Then the probability of the event ``$X\neq Y$'', i.e.
    $$P_e=1-\sum_j \tr(\rho X_jY_j),$$
    satisfies
    $$H(\alg{X}|\alg{Y})\leq 1+P_e\log\tr\1_{\alg{X}}.$$
  \end{lemma}
  \begin{proof}
    See~\cite{winter:qdms}, corollary~A.25. Observe that the statement
    of the lemma is a way of expressing the Holevo bound~\cite{holevo:bound}.
  \end{proof}

\section{Upper Capacity Bounds}
  \label{sec:upperbounds}
  The following theorem (which we call the weak converse because
  of theorem~\ref{satz:direct:coding} and note~\ref{bem:ahlswede:cap})
  was, in the case $r=1$ and $s=2$, stated in~\cite{allahverdyan:saakian}.
  \begin{satz}[Weak converse]
    \label{satz:weak:converse}
    The capacity region of the quantum multiway channel is contained in the
    closure of all nonnegative
    $(R_1,\ldots,R_s)$ satisfying for all $J\subset[s]$ and $j\in[r]$
    $$R(J)=\sum_{i\in J} R_i
        \leq\sum_u q_u I_{\gamma_u}\left(\alg{X}(J)\wedge\alg{Y}_j|\alg{X}(J^c)\right)$$
    for channel states $\gamma_u$ and $q_u\geq 0$, $\sum_u q_u=1$.
  \end{satz}
  \begin{proof}
    Let $(f_1,\ldots,f_s,D_1,\ldots,D_r)$ be any
    $(n,\bar{\epsilon})$--code with rate tuple $(R_1,\ldots,R_s)$.
    Then the uniform distribution on the codewords induces a channel state
    $\gamma$ on $(\alg{X}_1\cdots\alg{X}_s\alg{Y})^{\otimes n}$:
    \[\gamma=\frac{1}{|\fset{M}_1|\cdots|\fset{M}_s|}
             \sum_{\forall i:\ m_i\in\fset{M}_i}
                 \bigotimes_i f_i(m_i)\otimes W^n_{f_1(m_1)\ldots f_s(m_s)}.\]
    Its restriction
    to the $u$--th copy in this tensor power will be denoted $\gamma_u$.
    Let $j\in[r]$, $J\subset[s]$:
    by Fano inequality (lemma~\ref{lemma:fano}) we have
    \[H(\alg{X}^{\otimes n}(J)|\alg{Y}^{\otimes n}_j\alg{X}^{\otimes n}(J^c))
                                                  \leq 1+\bar{\epsilon}\cdot nR(J).\]
    With
    \begin{equation*}\begin{split}
      H&(\alg{X}^{\otimes n}(J)|\alg{Y}^{\otimes n}_j\alg{X}^{\otimes n}(J^c)) \\
       &\phantom{=====}
        =H(\alg{X}^{\otimes n}(J))
         -I(\alg{X}^{\otimes n}(J)\wedge\alg{Y}^{\otimes n}_j\alg{X}^{\otimes n}(J^c)) \\
       &\phantom{=====}
        =nR(J)
         -I(\alg{X}^{\otimes n}(J)\wedge\alg{Y}^{\otimes n}_j\alg{X}^{\otimes n}(J^c))
    \end{split}\end{equation*}
    we conclude now
    \begin{equation*}\begin{split}
      (1-\bar{\epsilon})R(J)
              &\leq \frac{1}{n}+\frac{1}{n}
                     I_\gamma(\alg{X}^{\otimes n}(J)\wedge
                               \alg{Y}^{\otimes n}_j\alg{X}^{\otimes n}(J^c)) \\
              &\leq \frac{1}{n}+\frac{1}{n}\sum_{u=1}^{n}
                     I_{\gamma_u}(\alg{X}(J)\wedge\alg{Y}_j\alg{X}(J^c)),
    \end{split}\end{equation*}
    using lemma 1 (subadditivity of mutual information).
  \end{proof}
  \begin{bem}
    \label{bem:ahlswede:cap}
    For classical channels the region described in the theorem is the exact capacity
    region (i.e. all the rates there are achievable), as was first proved by
    Ahlswede~\cite{ahlswede:MAC,ahlswede:MWC}. This fact is our reason to call
    it the \emph{weak converse}, as it describes the best outer bounds
    of $(n,\bar{\epsilon})$--code rates for $n\rightarrow\infty$
    and $\bar{\epsilon}\rightarrow 0$.
    \par
    To prove that for multiple access channels ($r=1$) this holds, too,
    is the object of the rest of the paper, though we conjecture it to
    be true in general.
  \end{bem}
  \begin{bem}
    The numerical computation of the above regions is not yet possible from
    the given description: we need a bound on the number of different single--letter
    channel states one has to consider in the convex combinations.
    For the multiple access channel ($r=1$) this is easy: by Caratheodory's theorem
    $s$ will suffice. For general $r$ it is possible to show
    that $r(2^s-1)$ are sufficient (cf.~\cite{csiszar:koerner}).
  \end{bem}

\section{Measurement Error and Disturbance}
  \label{sec:distortion:fidelity}
  \label{sec:decoding:operation}
  In this section a central result is proved that essentially
  states that if a POVM serves to indentify the states of an ensemble
  with high probability, then it may be implemented as an
  operation that disturbes the ensemble states very little.
  \begin{lemma}
    \label{lemma:measure:operation}
    Let $\rho$ be a state, and $X$ a positive operator with $X\leq\1$
    and $1-\tr(\rho X)\leq\epsilon<1$. Then
    $$\|\rho-\sqrt{X}\rho\sqrt{X}\|_1\leq\sqrt{8\epsilon},$$
    with the trace norm $\|X\|_1=\tr|X|$.
  \end{lemma}
  \begin{proof}
    See~\cite{winter:qstrong}, lemma~V.9.
  \end{proof}
  \begin{lemma}[Tender measurement]
    \label{lemma:tender:measurement}
    \label{lemma:tender:reconstruction}
    Let $\rho_a$ ($a\in\fset{A}$) be a set of states on $\alg{A}$,
    and $D$ an observable
    indexed by $\fset{B}$. Let further $\varphi:\fset{A}\rightarrow\fset{B}$ be any
    map and $\epsilon>0$ such that
    \begin{equation}
      \label{eq:tender}
      \forall a\in\fset{A}\qquad 1-\tr(\rho_a D_{\varphi(a)})\leq \epsilon
    \end{equation}
    (i.e. the observable recognizes $\varphi(a)$ from $\rho_a$ with maximal error
    probability $\epsilon$).
    Then the quantum operation $\delta:\alg{S}(\alg{A})\rightarrow\alg{S}(\alg{A})$
    defined by
    $$\delta:\rho\mapsto \sum_{b\in\fset{B}}\sqrt{{D}_b}\rho\sqrt{{D}_b}$$
    disturbes the states $\rho_a$ only a little:
    $$\forall a\in\fset{A}\qquad 
            \|\rho_a-\delta(\rho_a)\|_1\leq \sqrt{8\epsilon}+\epsilon.$$
    The quantum operation
    $\Delta:\alg{S}(\alg{A})\rightarrow\alg{S}(\C\fset{B}\otimes\alg{A})$
    with
    $$\Delta:\rho\mapsto 
                       \sum_{b\in\fset{B}} b\otimes\sqrt{{D}_b}\rho\sqrt{{D}_b}$$
    has the property that
    $$\forall a\in\fset{A}\qquad
              \|\varphi(a)\otimes\rho_a-\Delta(\rho_a)\|_1
              \leq \sqrt{8\epsilon}+\epsilon.$$
  \end{lemma}
  \begin{proof}
    It suffices to prove the second statement since the first
    inequality is obtained from it
    by a partial trace which does not increase $\|\cdot\|_1$:
    \begin{equation*}\begin{split}
      \left\|\varphi(a)\otimes\rho_a-\Delta(\rho_a)\right\|_1
        &\leq \left\|\rho_a-\sqrt{{D}_{\varphi(a)}}\rho_a\sqrt{{D}_{\varphi(a)}}\right\|_1\\
        &\phantom{====}+\sum_{b\neq \varphi(a)} \|\sqrt{{D}_b}\rho_a\sqrt{{D}_b}\|_1       \\
        &=    \left\|\rho_a-\sqrt{{D}_{\varphi(a)}}\rho_a\sqrt{{D}_{\varphi(a)}}\right\|_1\\
        &\phantom{====}+\sum_{b\neq \varphi(a)} \tr(\rho_a{D}_b)                           \\
        &\leq \sqrt{8(1-\tr(\rho_a{D}_{\varphi(a)}))}                                     \\
        &\phantom{====}+(1-\tr(\rho_a{D}_{\varphi(a)}))                                    \\
        &\leq \sqrt{8\epsilon}+\epsilon,
    \end{split}\end{equation*}
    using triangle inequality and lemma~\ref{lemma:measure:operation}.
  \end{proof}
  \begin{lemma}[Average version]
    \label{lemma:average:tenderness}
    With the same definitions as in lemma~\ref{lemma:tender:measurement},
    only replacing equation (\ref{eq:tender}) with
    \begin{equation}
      \label{eq:average:tender}
      \sum_{a\in\fset{A}} P(a)\left(1-\tr(\rho_a D_{\varphi(a)})\right)\leq\bar{\epsilon},
    \end{equation}
    for a probability distribution $P$ on $\fset{A}$, we have
    $$\sum_{a\in\fset{A}} P(a)\|\rho_a-\delta(\rho_a)\|_1
                  \leq \sqrt{8\bar{\epsilon}}+\bar{\epsilon}$$
    and
    $$\sum_{a\in\fset{A}} P(a)\|\varphi(a)\otimes\rho_a-\Delta(\rho_a)\|_1
                  \leq \sqrt{8\bar{\epsilon}}+\bar{\epsilon}.$$
  \end{lemma}
  \begin{proof}
    Again, we have only to prove the second statement.
    Introducing $\epsilon_a=1-\tr(\rho_a D_{\varphi(a)})$
    we have, like in the previous proof,
    $$\left\|\varphi(a)\otimes\rho_a-\Delta(\rho_a)\right\|_1
           \leq \sqrt{8\epsilon_a}+\epsilon_a.$$
    Forming the average of the left hand side under the distribution $P$,
    and using concavity of $\sqrt{8x}+x$ the assertion follows.
  \end{proof}

\section{Quantum Multiple Access Channel: Coding}
  \label{sec:directcoding}
  Throughout this section we will assume $r=1$ and else notation as before.
  \begin{satz}
    \label{satz:direct:coding}
    Let $R_1,\ldots,R_s$ be nonnegative, satisfying for some a priori distributions
    $P_i$ on the $\fset{X}_i$ the constraints
    \[\forall J\subset[s]\qquad
          \sum_{i\in J} R_i\leq I\left(\alg{X}(J)\wedge\alg{Y}|\alg{X}(J^c)\right).\]
    Then for every $\bar{\epsilon},\delta>0$ and all sufficiently large
    $n$ there are $(n,\bar{\epsilon})$--codes with rates
    $\frac{1}{n}\log|\fset{M}_i|\geq R_i-\delta$.
  \end{satz}
  \begin{proof}
    It is sufficient by the time sharing principle
    to prove the assertion only for
    the upper extremal points of the region described, and by symmetry we may assume
    (for $i=1,\ldots,s$) that
    \begin{equation*}\begin{split}
      R_i &= I(\alg{X}_i\wedge\alg{Y}\alg{X}_1\cdots\alg{X}_{i-1}) \\
          &= H(\alg{Y}|\alg{X}_1\cdots\alg{X}_{i-1})
            -H(\alg{Y}|\alg{X}_1\cdots\alg{X}_i)                   \\
          &= H(P_{\{>i-1\}}W|P_{\{\leq i-1\}})
            -H(P_{\{>i\}}W|P_{\{\leq i\}}).
    \end{split}\end{equation*}
    That these are indeed the upper extreme points is proved in
    the appendix.
    \par
    The idea of the following construction is to first decode the message
    $m_1$ from sender $1$, using only the incoming signal.
    Then decode the message $m_2$ from sender $2$, using $m_1$ and the incoming
    signal (which is almost undisturbed by the tender measurement lemma).
    Iterate, until you decode message $m_s$ from sender $s$, using
    $m_1,\ldots,m_{s-1}$ and the still almost unchanged incoming signal.
    \par
    Let $\epsilon,\delta>0$, and
    consider $s$ families of codewords $\fset{C}_i\subset\fset{X}_i^n$ of size
    $L_i=|\fset{C}_i|=\lceil 2^{n(R_i-\delta)}\rceil$,
    drawn independently from $\fset{X}_i^n$
    according to the a priori distribution $P_i^{\otimes n}$.
    \par
    Fix $i$ for the moment and define the following channel:
    for $x_i^n\in\fset{X}_i^n$
    \begin{equation*}\begin{split}
      \rho^{(i)} &:x_i^n \mapsto \rho_{x_i^n}^{(i)}= \\
                 &\!\!\!\!
                  \frac{1}{L_1\cdots L_{i-1}L_{i+1}\cdots L_s}
                   \sum_{\forall j\neq i:\ c_j^n\in\fset{C}_j}\!\!
                   \left(\!\bigotimes_{j<i} c_j^n\otimes
                           W^n_{c_1^n\ldots x^n_i\ldots c_s^n}\!\right)
    \end{split}\end{equation*}
    (we denote these word states by $\rho^{(i)}$, in contrast to the letter states $W$).
    Note that this is the channel belonging to the channel
    state $\gamma$ from the proof of theorem~\ref{satz:weak:converse},
    reduced to $(\alg{X}_1\cdots\alg{X}_{i-1}\alg{Y})^{\otimes n}$.
    \par
    The average of $\rho^{(i)}_{x_i^n}$ over the choice of
    $\fset{C}_1,\ldots,\fset{C}_{i-1}$, $\fset{C}_{i+1},\ldots,\fset{C}_s$
    is indeed a product state:
    \begin{equation*}\begin{split}
      \langle\rho^{(i)}_{x_i^n}\rangle_{\fset{C}_i:i\neq j}
                 &=\!\!\sum_{\forall j\neq i:\ x_j^n\in \fset{X}^n_j}\!\!
                   P_{\{\neq i\}}(x_j^n|j\neq i)\bigotimes_{j<i} x_j^n\otimes
                   W^n_{x_1^n\ldots x_s^n}                               \\
                 &\hspace{-1.2cm}
                  =\!\!\sum_{\forall j<i:\ x_j^n\in \fset{X}^n_j}\!\!
                   P_{\{<i\}}(x_j^n|j<i)\bigotimes_{j<i} x_j^n\otimes
                   (P_{\{>i\}}W)^n_{x_1^n\ldots x_i^n}                   \\
                 &\hspace{-1.2cm}=V^n_{x_i^n},
    \end{split}\end{equation*}
    with
    $$V_{x_i}=\!\!\sum_{\forall j<i:\ x_j\in\fset{X}_j}\!\!
               P_{\{<i\}}(x_j|j<i)\bigotimes_{j<i} x_j
                                \otimes(P_{\{>i\}}W)_{x_1\ldots x_i}.$$
    In~\cite{holevo:capacity} and~\cite{schumacher:capacity}
    a construction of a decoding observable for the channel
    $V$ and set $\fset{C}_i$ of
    codewords is described,\footnote{Observe that in this way
      $D_i$ will be independent from the other codes
      and their decoding observables!}
    which has the property that it's average error probability
    $$\bar{e}_{V^n}(\fset{C}_i,D_i)=1-\frac{1}{|\fset{C}_i|}
                     \sum_{c_i^n\in\fset{C}_i} \tr(V^n_{c_i^n}D_{i,c_i^n}),$$
    averaged over the choice of $\fset{C}_i$, is at most
    $\epsilon/s$ for large enough $n$:
    $${\langle\bar{e}_{V^n}(\fset{C}_i,D_i)\rangle}_{{\cal C}_i}\leq\epsilon/s$$
    (where we identified the set of messages with $\fset{C}_i$).
    This is because
    $I(P_i;V)=I(\alg{X}_i\wedge\alg{Y}\alg{X}_1\cdots\alg{X}_{i-1})=R_i$.
    (Recall the approach of~\cite{holevo:capacity} and~\cite{schumacher:capacity}:
    a random code --- drawn according to $P_i^{\otimes n}$ --- is chosen
    with rate slightly below $I(P_i;V)$, and a decoding POVM constructed
    which forces the expected average error probability small. Then
    it is concluded that a code with this small error probability
    actually exists).
    \par
    We note that by the construction from~\cite{holevo:capacity}
    and~\cite{schumacher:capacity} it is assured that
    $$D_{i,c_i^n}\in(\alg{X}_1\cdots\alg{X}_{i-1}\alg{Y})^{\otimes n},$$
    for all $c_i^n\in\fset{C}_i$.
    It is easlity seen that we may assume this w.l.o.g., for
    wherever the $D_i$ comes from: the $V^n_{c_i^n}$ are density
    operators on some Hilber space ${\cal H}$, such that
    $$V^n_{c_i^n}\in(\alg{X}_1\cdots\alg{X}_{i-1}\alg{Y})^{\otimes n}
                 \subset\alg{L}({\cal H}).$$
    Denoting this subset embedding by $E$, we have
    $$\tr(V^n_{c_i^n}D_{i,c_i^n})
           =\tr(E(V^n_{c_i^n})D_{i,c_i^n})
           =\tr(V^n_{c_i^n}E^*(D_{i,c_i^n})),$$
    with the adjoint map $E^*$. I.e. the decoding error probabilities
    do not change if we replace the $D_{i,c_i^n}$ by
    $E^*(D_{i,c_i^n})\in(\alg{X}_1\cdots\alg{X}_{i-1}\alg{Y})^{\otimes n}$.
    \par
    Now we confront the decoder $D_i$ with the signal $\rho$, obtaining an average
    error probability
    $$\bar{e}_i=\bar{e}_{\rho^{(i)}}(\fset{C}_i,D_i)
               =1-\frac{1}{L_i}\sum_{c_i^n\in\fset{C}_i}
                                   \tr(\rho_{c_i^n}D_{i,c_i^n}).$$
    Obviously the code average over choice of $\fset{C}_j,\ j\neq i$ is exactly
    $\bar{e}_{V^n}(\fset{C}_i,D_i)$, so by the previous argument
    $${\langle\bar{e}_1+\ldots+\bar{e}_s\rangle}_{\fset{C}_1\ldots\fset{C}_s}
                                                                 \leq\epsilon.$$
    This means that there actually \emph{exist}
    codebooks $\fset{C}_1,\ldots,\fset{C}_s$
    with all the $\bar{e}_i\leq\epsilon$.
    Now consider the corresponding operations
    $\Delta_i$ of the observables $D_i$, from lemma~\ref{lemma:tender:measurement}.
    By the average tender measurement lemma~\ref{lemma:average:tenderness}
    we find for each $i$:
    \begin{equation*}\begin{split}
      &\frac{1}{L_1\!\cdots\! L_s}\!
       \sum_{\forall j:c_j^n\in\fset{C}_j}
           \!\left\|\bigotimes_{j\leq i} c_j^n \!\otimes\! W^n_{c^n_1\ldots c_s^n}
           \!-\!\Delta_i\!\!\left(\!\bigotimes_{j<i} c_j^n \!\otimes\! W^n_{c^n_1\ldots c_s^n}
                               \!\!\right)\!\right\|_1         \\
      &\phantom{====}\leq \sqrt{8\epsilon}+\epsilon.
    \end{split}\end{equation*}
    Their concatenation
    $\Delta=\Delta_s\circ\cdots\circ\Delta_1$
    will be the decoder process: it satisfies
    \begin{equation*}\begin{split}
      &\frac{1}{L_1\cdots L_s}\sum_{\forall j:c_j^n\in\fset{C}_j}
           \left\|\bigotimes_{j=1}^s c_j^n\otimes W^n_{c^n_1\ldots c_s^n}
                  -\Delta(W^n_{c^n_1\ldots c_s^n})\right\|_1       \\
      &\phantom{===}\leq s(\sqrt{8\epsilon}+\epsilon),
    \end{split}\end{equation*}
    since each step introduces a deviation of at most
    $\sqrt{8\epsilon}+\epsilon$, and
    does not increase trace norm.
    \par
    Now, discarding the signal itself (i.e tracing out $\alg{Y}^{\otimes n}$),
    we arrive at $\widetilde{\Delta}=\tr_{\alg{Y}^{\otimes n}}\circ\Delta$,
    which satisfies
    \begin{equation*}
      \frac{1}{L_1\cdots L_s}\sum_{\forall j:c_j^n\in\fset{C}_j}
           \left\|\bigotimes_{j=1}^s c_j^n
                  -\widetilde{\Delta}(W^n_{c^n_1\ldots c_s^n})\right\|_1
           \leq s(\sqrt{8\epsilon}+\epsilon).
    \end{equation*}
    This means, that $\widetilde{\Delta}$ is a quantum operation
    with classical outcomes in $\fset{C}_1\times\cdots\times\fset{C}_s$
    (hence described by a POVM $D$ indexed by this set),
    which recovers rather accurately $(c^n_1,\ldots,c_s^n)$
    from $W^n_{c^n_1\ldots c_s^n}$, on average.
  \end{proof}
  \par\medskip
  Now we may combine the (weak) converse
  theorem~\ref{satz:weak:converse}
  and the foregoing direct coding theorem~\ref{satz:direct:coding}
  (together with the time sharing principle
  by which the capacity region is convex) to obtain
  \begin{satz}
    \label{satz:mac:capacity}
    The capacity region of the quantum multiple access channel is the
    convex closure of all nonnegative
    $(R_1,\ldots,R_s)$ satisfying
    \[\forall J\subset[s]\qquad \sum_{i\in J} R_i\leq
                     I\left(\alg{X}(J)\wedge\alg{Y}|\alg{X}(J^c)\right)\]
    for some input distribution and corresponding channel state.
    \qed
  \end{satz}

\section{Quantum--Quantum Multiway Channels}
  \label{sec:qq:dmc}
  Allahverdyan and Saakian in~\cite{allahverdyan:saakian} actually defined
  quantum multiway channels in more general form: namely allowing in the
  definition of section~\ref{sec:channel} general C${}^*$--algebras
  $\alg{X}_i$ (and not only commutative ones as we did).
  Each sender $i$ may then use any set
  of states on $\alg{X}_i$ for transmission. Formally, let
  $$\varphi:\alg{S}(\alg{X}_{1}\otimes\cdots\otimes\alg{X}_{s})
                                     \rightarrow\alg{S}(\alg{Y})$$
  be any completely positive, trace preserving, linear map,
  i.e. a quantum operation on states.
  A code for sender $i$ now consists of a map
  $F_i:\fset{M}_i\rightarrow\alg{S}(\alg{X}^{\otimes n}_i)$,
  the decoders are still observables in the algebras $\alg{Y}_j$.
  This allows for definition of error probabilities and rates,
  and hence of the capacity region, the details of which we leave to
  the reader.
  \par
  By composition we can view this as a classical--quantum
  channel on $n$--blocks:
  $$(m_1,\ldots,m_s)\mapsto
                   \varphi^{\otimes n}(F_1(m_1)\otimes\cdots\otimes F_s(m_s)).$$
  If we now make the restriction
  that in coding the senders have to use \emph{product states},
  i.e.
  $$F_i(m_i)=F_{i,m_i,1}\otimes\cdots\otimes F_{i,m_i,n}$$
  with $F_{i,m_i,k}\in\alg{S}(\alg{X}_i)$,
  we see immediately that the argument from section~\ref{sec:upperbounds}
  applies, and gives a general upper bound theorem
  (with the slight difference that now in the formation of the channel
  states $\gamma_u$ instead of the states $W_{f_1(m_1)_u\ldots f_s(m_s)_u}$
  there occur the output states
  $\varphi(F_{1,m_1,u}\otimes\cdots\otimes F_{s,m_s,u})$).
  \par
  Conversely: those input states and a distribution on them being chosen we
  can apply the coding theorem~\ref{satz:direct:coding}
  to see that these bounds can be approached --- at least for the multiple
  access channel.
  \par
  We have thus obtained the capacity region ${\bf R}^{(1)}$
  in the \emph{1--separable} case,
  i.e. for input states that are separable (it is easy to see that
  best performance in this case
  is for product states). Of course we could allow the sender to
  use blocks of states which entangle at most $\nu$
  systems, getting a possibly larger
  region ${\bf R}^{(\nu)}$. This is because now the subadditivity
  of the mutual information, used in the upper bound
  in section~\ref{sec:upperbounds}, does not hold any longer.
  We leave open the question of the ultimate capacity region
  ${\bf R}=\overline{\bigcup_\nu {\bf R}^{(\nu)}}$
  (which even for the case $s=r=1$ is unsolved):
  whether it be equal or strictly larger than ${\bf R}^{(1)}$.

\section{Conclusion}
  \label{sec:conclusion}
  We introduced quantum multiway channel and bounded its capacity region.
  In the case of the quantum multiple access channel this led to a complete
  characterization of the capacity region,
  formally identical to the result in the classical case. This confirms the validity
  of the principle that in the theory of classical information transmission
  over a quantum channel capacities are
  obtained from the classical formulas by replacing Shannon entropy by
  von Neumann entropy.
  \par
  The next step is of course to prove the direct coding theorem in the case of
  several receivers. It seems that for this a refinement of the methods used here
  or even a different approach is needed: namely, it is known that
  it is not sufficient to achieve rate tuples of the form used in the
  proof of theorem~\ref{satz:direct:coding}. Hence the method of sequentially
  decoding the messages of the different senders has to be abandoned.
  \par\bigskip
  {\bf A note on the literature}:
  Since the findings of the present paper (published at that time
  only as an e--print ({\tt quant-ph/9807019}))
  an article by Huang, Zhang, and
  Hou~\cite{hzh} on quantum multiple access channels has appeared,
  where the 2--sender pure state case of a multiple access channel
  is treated.
  \par
  In that work a reference to the above statement was made concerning
  the wish for a more direct proof of the achievability of the
  region described in the coding theorem~\ref{satz:mac:capacity}.
  The authors propose to supply such a proof.
  However, they too employ the
  present paper's method to show achievability of the
  ``corners'' (compare theorem~\ref{satz:direct:coding}),
  and then invoke the
  time--sharing principle. As was pointed out here this method
  is not adapted to prove the theorem for multiple receivers:
  it is still an open problem to find a proof that
  directly allows to achieve every point of the capacity
  region. We most probably know how to perform the random
  code selection, but the analog of the maximum likelihood
  decoder necessary to decode all messages simultaneously
  still evades us (for detail how this
  strategy works in the classical case,
  see~\cite{ahlswede:MWC}).

\section*{Acknowledgements}
  The author wishes to thank Peter L\"ober for discussions
  on the subject, and Prof. Rudolf Ahlswede
  for comments on the proof of the classical theorem.

\appendix
\section*{Extremal points}
  \label{sec:extremal}
  We shall prove: any upper extremal point of the region of
  non--negative tuples $(R_1,\ldots,R_s)$
  which satisfy for all $J\subset[s]$
  \begin{equation}
    \label{eq:J}
    R(J)\leq I(\alg{X}(J)\wedge\alg{Y}\alg{X}(J^c))\tag{{\it J}},
  \end{equation}
  for fixed channel state, is one of the $(R_1,\ldots,R_s)$ with
  $$R_{\pi(i)}=
     I(\alg{X}_{\pi(i)}\wedge\alg{Y}\alg{X}_{\pi(1)}\cdots\alg{X}_{\pi(i-1)})$$
  for some permutation $\pi$ of the set $[s]$, and any such tuple belongs to this region.
  \par
  Let $(R_1,\ldots,R_s)$ be an upper extremal point, hence $s$
  of the inequalities (for
  nonempty subsets $K_1,\ldots,K_s$ of $[s]$) are met with
  equality. We are done if we can prove that these can be chosen to form a
  strictly increasing chain
  $\emptyset\neq K_1\subset K_2\subset\ldots\subset K_s=[s]$, w.l.o.g.
  $K_i=\{s,s-1,\ldots,s-i+1\}$,
  because then
  \begin{equation}\begin{split}
    \label{eq:extrem}
    R_i &=I(\alg{X}_i\cdots\alg{X}_s\wedge\alg{Y}\alg{X}_1\cdots\alg{X}_{i-1}) \\
        &\phantom{==}
         -I(\alg{X}_{i+1}\cdots\alg{X}_s\wedge\alg{Y}\alg{X}_1\cdots\alg{X}_i) \\
        &=I(\alg{X}_i\wedge\alg{Y}\alg{X}_1\cdots\alg{X}_{i-1}).
  \end{split}\end{equation}
  Choose any one such set, say $K\neq\emptyset$:
  $$R(K)=I(\alg{X}(K)\wedge\alg{Y}\alg{X}(K^c)).$$
  We would like to employ induction to extend this $K$ in both directions
  to a chain. That we can do so follows from
  \begin{lemma}
    \label{lemma:extremal}
    If $R_1,\ldots,R_s\geq 0$ are such that with $\emptyset\neq K\subset[s]$:
    $R(K)=I(\alg{X}(K)\wedge\alg{Y}\alg{X}(K^c))$ and $(J)$ holds for all
    $J\subset K$ and for all $K\subset J\subset[s]$.
    Then $(J)$ holds for all $J\subset[s]$.
  \end{lemma}
  \begin{proof}
    First let $K\subset J\subset[s]$.
    Then $(J)$, together with $(K)$ met with equality,
    implies
    \begin{equation*}\begin{split}
      R(J\setminus K)
           &\leq H(\alg{Y}|\alg{X}(J^c))-H(\alg{Y}|\alg{X}(K^c))              \\
           &\leq H(\alg{Y}|\alg{X}((J\setminus K)^c))-H(\alg{Y}|\alg{X}([s])) \\
           &= I(\alg{X}(J\setminus K)\wedge\alg{Y}|\alg{X}((J\setminus K)^c)),
    \end{split}\end{equation*}
    which is equation $(J\setminus K)$.
    There the second estimate is by strong subadditivity applied to
    $\alg{A}_1=\alg{X}(J\setminus K)$,
    $\alg{A}_2=\alg{Y}\alg{X}(J^c)$, $\alg{A}_3=\alg{X}(K)$.
    \par
    Now let $J\subset[s]$ be arbitrary and write $J=J_1\dot{\cup}J_2$ with
    $J_1=J\cap K$, $J_2=J\setminus K$. Then by assumption and the previous
    \begin{equation*}\begin{split}
      R(J) &=     R(J_1)+R(J_2)                                             \\
           &\leq  H(\alg{Y}|\alg{X}(J_1^c))-H(\alg{Y}|\alg{X}([s]))         \\
           &\qquad      +H(\alg{Y}|\alg{X}(J_2^c\cap K^c))-H(\alg{Y}|\alg{X}(K^c)) \\
           &=     H(\alg{Y}\alg{X}(J_1^c))+H(\alg{Y}\alg{X}(K^c\cap J_2^c))
                 -H(\alg{Y}\alg{X}(K^c))                                    \\
           &\qquad   -H(\alg{X}((K\setminus J_1)\dot{\cup}(K^c\setminus J_2)))
                     -H(\alg{Y}|\alg{X}([s]))                               \\
           &\leq  H(\alg{Y}\alg{X}(J_1^c\cap J_2^c))-H(\alg{X}(J_1^c\cap J_2^c))
                 -H(\alg{Y}|\alg{X}([s]))                                   \\
           &=     H(\alg{Y}|\alg{X}(J_1^c\cap J_2^c))-H(\alg{Y}|\alg{X}([s]))\\
           &=     I(\alg{X}(J)\wedge\alg{Y}|\alg{X}(J^c)).
    \end{split}\end{equation*}
    There the second estimate is again by strong subadditivity, with
    $\alg{A}_1=\alg{X}(J_2)$, $\alg{A}_2=\alg{Y}\alg{X}(K^c\cap J_2^c)$,
    $\alg{A}_3=\alg{X}(K\setminus J_1)$.
  \end{proof}
  \par\medskip
  It follows that if $(K')$ is met with equality for any $K'$, also
  $(K'\cap K)$ and $(K'\cup K)$ are met with equality. It is easily seen that
  both below and above $K$ there must occur equalities (except
  already $|K|=1$ or $K=[s]$).
  \par
  It remains to show that $(R_1,\ldots,R_s)$ defined by (\ref{eq:extrem})
  belongs to the region. But this is clear by the same argument as above, used
  top--down: the conjunction of all $(J)$ with $J\subset[s]$ is implied by the
  conjunction of all with $J\subset[s-1]$, this in turn inductively
  by all with $J\subset[s-2]$ etc.

\end{document}